\documentclass[%
reprint,
superscriptaddress,
 amsmath,amssymb,
 aps,
 prb,
]{revtex4-2}

\allowdisplaybreaks

\usepackage{graphicx}
\usepackage{dcolumn}
\usepackage{bm}

\usepackage{algorithm}
\usepackage{algpseudocode}
\usepackage{orcidlink}
\usepackage{float}

\begin{document}

\preprint{APS/123-QED}

\title{Variational Evolutionary Network for Statistical Physics Systems}
%
\author{Yixiong Ren~\orcidlink{0009-0000-9544-2888}}
\affiliation{Anhui Provincial Key Laboratory of Low-Energy Quantum Materials and Devices,
High Magnetic Field Laboratory, HFIPS, Chinese Academy of Sciences, Hefei, Anhui 230031, China.}
\affiliation{University of Science and Technology of China, Hefei 230026, P. R. China\index{ss}}

\author{Jianhui Zhou}
\email{jhzhou@hmfl.ac.cn}
\affiliation{Anhui Provincial Key Laboratory of Low-Energy Quantum Materials and Devices,
High Magnetic Field Laboratory, HFIPS, Chinese Academy of Sciences, Hefei, Anhui 230031, China.}
\date{\today}

%
\begin{abstract}
Monte Carlo methods are widely used importance sampling techniques for studying complex physical systems. Integrating these methods with deep learning has significantly improved efficiency and accuracy in high-dimensional problems and complex system simulations. However, these neural network-enhanced Monte Carlo methods still face challenges such as slow sampling speeds, statistical bias, and inaccuracies in the ground state. To address these issues, we propose a variational evolutionary network, which utilizes neural networks for variational free energy and combines evolutionary algorithms for sampling. During the sampling process, we construct generation and selection operators to filter samples based on importance, thereby achieving efficient importance sampling. We demonstrate that this sampling method provides an upper bound on the ground-state energy, enhancing both sampling efficiency and ground-state accuracy. Moreover, we numerically examine our method in two-dimensional Ising model and Sherrington-Kirkpatrick Model for spin glass. Thus, our algorithm could offer improved accuracy in handling complex energy landscapes and significantly enhance computational efficiency.

\end{abstract}

\maketitle


\section{\label{sec:level1}INTRODUCTION}

Monte Carlo simulations are widely used for studying complex physical systems, especially for analyzing thermodynamic properties, phase transitions, and critical phenomena~\cite{binder_monte_2010, krauth_statistical_2006}. These simulations use random sampling to handle systems where obtaining analytical solutions is difficult. However, they face challenges such as slow convergence, low computational efficiency, and getting trapped in local minima when dealing with high-dimensional systems, states near critical points of phase transitions, and complex energy landscapes~\cite{wolff_collective_1989, swendsen_nonuniversal_1987, hukushima_exchange_1996}. Various improved algorithms have been proposed to overcome these limitations, including the Wolff algorithm~\cite{wolff_collective_1989}, the Swendsen-Wang algorithm~\cite{swendsen_nonuniversal_1987}, and parallel tempering Monte Carlo methods~\cite{hukushima_exchange_1996}. While these algorithms have made significant progress in enhancing simulation efficiency and expanding the applicable range, they are typically designed for specific problems or systems and are not easily generalizable to all types of physical systems.

With the development of neural networks, Wu proposed a new algorithm called the Variational Autoregressive Networks (VAN)~\cite{wu_solving_2019}, in which variational methods are used to train a neural network to model the target distribution and an autoregressive approach is for sampling. Unlike traditional Monte Carlo methods, VAN does not rely on Markov chains to ensure sampling converges to the equilibrium distribution. Instead, it utilizes a free energy variational approach to simulate the conditional probability distribution of the model. This method has shown significant advantages in addressing the problem of critical slowing down and has a broader applicability. However, the VAN method still has limitations, such as biased estimates, slow training speed, and inaccurate ground state sampling when one deals with frustrated spin models~\cite{ciarella_machine-learning-assisted_2023, bialas_analysis_2023}.

Several improvements have been proposed to overcome the issues of biased estimates and slow training speeds associated with the VAN method. Some approaches involve computing sample weights to reduce bias~\cite{nicoli_asymptotically_2020, mcnaughton_boosting_2020}, while others utilize neural networks to assist Monte Carlo sampling or enhance sampling diversity through model symmetry~\cite{wu_unbiased_2021}. The problem of slow training speed primarily arises from the autoregressive sampling method, where each sampling iteration for a model with $N$ lattice points requires $N$ neural network inferences, leading to increase of computational costs. Białas~\cite{bialas_hierarchical_2022} partitioned the model into multiple segments to improve sampling efficiency, allowing for hierarchical sampling in which neural networks can sample each segment in parallel, thereby reducing the overall computational time.

Beyond issues of biased estimates and slow training, the complex energy landscape of the ground state of spin glass poses additional challenges for optimization algorithms, which usually get trapped in local minima. The VAN algorithm similarly struggles with these difficulties. Optimizing the neural network architecture has been focused on enhancing the accuracy of solutions~\cite{ma_message_2024}. Additionally, in computer science, many combinatorial optimization problems can be represented by spin glass models~\cite{lucas_ising_2014}, leading to the development of algorithms similar to VAN, such as the Gumbel-softmax based on Gumbel sampling~\cite{li_gumbel-softmax-based_2021, liu_gumbel-softmax_2019} and Monte Carlo Policy Gradient Method~\cite{chen_monte_2023}.

In this work, we propose a novel algorithm, the variational evolutionary network (VEN), inspired by the observed parallels between sampling from a target distribution to determine ground states and the iterative process of optimizing populations within evolutionary algorithms~\cite{slowik_evolutionary_2020} to achieve optimal solutions. 
In VEN, the samples generated can be viewed as a population within an evolutionary framework, where extensive sampling helps to reduce bias, akin to increasing diversity and coverage by expanding the population in evolutionary algorithms. Moreover, VEN could bypass the autoregressive sampling method and directly sample from the target distribution, thereby enhancing computational efficiency by avoiding the need for multiple neural network inferences per sample. 
We apply our algorithm to the two-dimensional $($2D$)$ Ising model and Sherrington-Kirkpatrick Model with frustration. 
%

The rest of the work is organized as follows: an introduction to the algorithm is in Section \ref{sec:Algorithms}, theoretical analysis is in Section \ref{sec:Upper}, numerical results are present in Section \ref{sec:Num}, and the conclusion and discussions are given in Section \ref{sec:Conclusions}.

\section{Algorithms}
\label{sec:Algorithms}
The algorithm is divided into two parts: the variational framework and the sampling method. In the variational framework, we describe the variational approach and compare it with some concepts in reinforcement learning. In the sampling method, we introduce a new sampling approach derived from evolutionary algorithms. The algorithm flowchart is shown in Fig.~\ref{fig1}, and the pseudocode is provided in Alg.~\ref{al}.
\subsection{variational framework}
In statistical physics models with $N$ interacting components, the joint probability distribution of the system’s state $\mathbf{s}$ follows the Boltzmann distribution: 
\begin{align}
p(\mathbf{s}) = \frac{e^{-\beta E(\mathbf{s})}}{Z},
\end{align}
where $\beta = 1/T$ is the inverse temperature, and $Z$ is the partition function. The computational complexity of calculating this joint probability increases exponentially with the number of lattice sites $N$, rendering direct computation infeasible for large systems. Consequently, an efficient approximation method is necessary to estimate these probabilities. Neural networks have proven to be highly effective in modeling complex probability distributions. Therefore, we choose a neural network to approximate the joint probability distribution, denoted as $p_{\theta}(\mathbf{s})$, where $\theta$ represents the network parameters. For each lattice site in the model, the marginal probability $p_\theta(s_i)$ is output by the neural network, and the joint probability is approximated as the product of these marginal probabilities, 
\begin{align}
p_{\theta}(\mathbf{s}) = \prod_i p_\theta(s_i). 
\end{align}
According to the second law of thermodynamics, the free energy $ F = U - S/\beta $ reaches its minimum at equilibrium. Here, the internal energy $ U = \sum_{x} p(x) E(x) $ tends to favor configurations with lower energy, implying that the modeled distribution assigns higher probabilities to lower-energy states. Meanwhile, the entropy $ S = -\sum_{x} p(x) \ln p(x) $ tends to maximize, encouraging a more uniform distribution across different configurations. To achieve the minimization of free energy at equilibrium, we employ a variational approach that optimizes the neural network parameters $ \theta $, making the modeled free energy $ F_\theta(\beta, s) $ approximate the actual free energy $ F(\beta) $:
\begin{equation}
\begin{aligned}
& F_\theta(\beta, s) - F(\beta) \\
= & \sum_{x} b(x)E(x) + \frac{1}{\beta} \sum_x b(x) \ln b(x) + \frac{1}{\beta} \ln Z \\
= & \frac{1}{\beta} \sum_x b(x) \ln \frac{b(x)}{p(x)} \\
= & \frac{1}{\beta} D_{\text{KL}}(b(x) \| p(x)).
\end{aligned}
\label{eq:KL}
\end{equation}
where the Kullback-Leibler (KL) divergence $D_{\text{KL}}(b(x) \| p(x))$, measures the discrepancy between the modeled distribution $ p_\theta $ and the Boltzmann distribution $ p $. By minimizing the KL divergence through variational free energy minimization, the neural network output distribution becomes closer to the true Boltzmann distribution.

The variational learning process can be understood through the framework of reinforcement learning~\cite{sutton_reinforcement_2018}$($ see Table.~\ref{tab:analogy} for the comparison $)$. In this analogy, the parameter $\beta$ represents the system's state, while the sample $s$ at a given state is treated as an action. The sampling method is viewed as a policy that governs how different actions are selected based on the state of the system. Free energy serves as a criterion for evaluating the value of a state, allowing us to define a state value function $F(\beta)$ that represents the free energy of state $\beta$. Similarly, we define a state-action value function $F_\theta(\beta, s)$, which is computed using the neural network's architecture and parameters $\theta$, to evaluate the value of action $s$ in state $\beta$. The advantage function, $A_\theta(\beta, s) = F_\theta(\beta, s) - F(\beta)$, is then introduced to measure the relative benefit of executing a particular action in a given state.

\begin{table*}
\caption{\label{tab:analogy}Analogy between Variational Learning and Reinforcement Learning}
\centering
\begin{tabular}{|c|c|}
\hline
\textbf{Variational Learning} & \textbf{Reinforcement Learning} \\ \hline
Parameter $\beta$ & State \\ \hline
Sample $s$ at a given state & Action \\ \hline
Sampling method & Policy \\ \hline
Free energy $F(\beta)$ & State value function~\cite{sutton_learning_1988} \\ \hline
State-action value function $F_\theta(\beta, s)$ & Action value function~\cite{sutton_learning_1988} \\ \hline
KL divergence $D_{\text{KL}}(b(x) \| p(x))$ & Advantage function~\cite{baird_advantage_1993} \\ \hline
\end{tabular}
\end{table*}

In the context of reinforcement learning, the advantage function $A_\theta(\beta, s)$ serves two primary purposes. First, it assesses the accuracy of the value estimates provided by the state-action value function and the state value function; when $A_\theta(\beta, s) \approx 0$, this indicates that $F_\theta(\beta,s)$ and $F(\beta)$ are providing accurate estimates of the value of states and actions. Second, it evaluates the relative performance of a specific action $s'$ in state $\beta$. If $A_\theta(\beta, s') > 0$, this implies that the action $s'$ has a higher value than the state's average value, making it a more preferable action with greater advantage. The goal of the policy is to increase the probability of selecting actions with higher advantages.

Similarly, in the variational framework, we optimize the neural network parameters $\theta$ by minimizing the KL divergence, aligning the modeled free energy closer to the true free energy. Given the difficulty of directly computing $F(\beta)$, we approximate it using the average of $F_\theta(\beta, s)$:
\begin{align}
F(\beta) \simeq \frac{1}{n}\sum_{i=1}^n F_\theta(\beta, s_i),
\end{align}
where $n$ is the batch size. This analogy with reinforcement learning helps us better understand the critical role of neural network structures and sampling methods in the variational learning framework for accurately modeling joint probabilities.

For statistical physics models, neural networks play a crucial role as feature extractors, significantly impacting the model's ability to compute free energy approximations efficiently. The architecture of a neural network determines its capacity to capture features and accurately estimate free energy effectively. Compared to traditional methods like the mean-field approximation and Bethe approximation, neural networks exhibit superior flexibility and nonlinear representational power, enabling them to identify complex patterns and dependencies. 

\subsection{sampling method}
\begin{figure*}
\includegraphics[width=\textwidth]{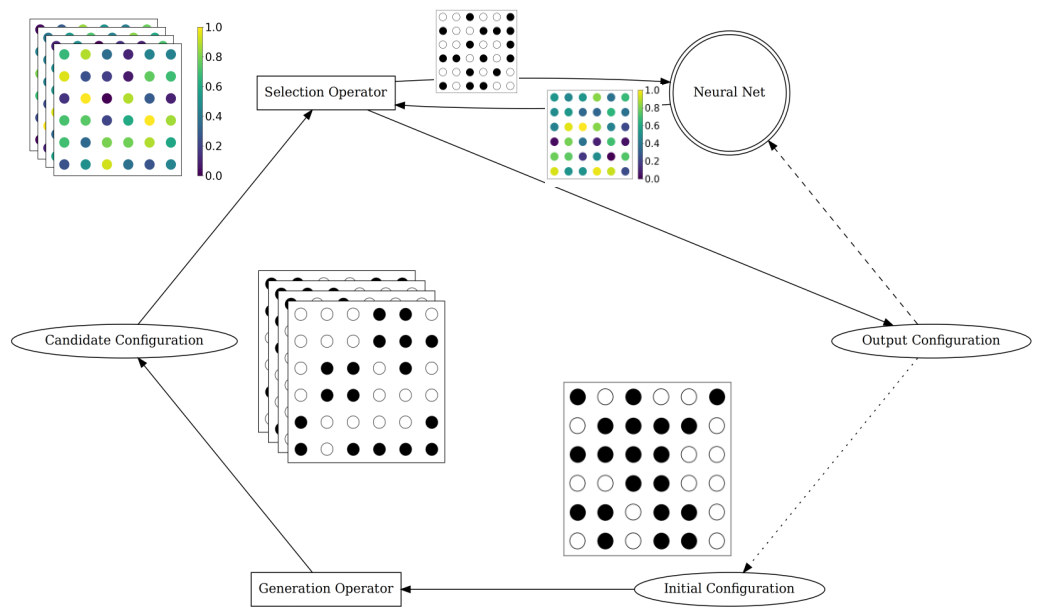}
\caption{\label{fig1} Structure Diagram of the VEN Sampling Method. Starting from an initial configuration, candidate configurations are first generated using a generation operator. Then, a selection operator utilizes a neural network to compute the probability distribution of the candidates, resulting in the output configuration. The output configuration is used both as the new initial configuration and for training the neural network through backpropagation.}
\end{figure*}

Sampling methods directly influence computational efficiency and variance. In the context of autoregressive sampling, known as variational autoregressive neural networks, the sampling process struggles to identify ground states in spin glass models. Observations indicate that, as the system transitions from high to low temperatures, the joint probability distribution progressively concentrates around lower energy configurations. This behavior is reminiscent of evolutionary algorithms, which gradually converge to an optimal solution through iterative processes that simulate natural selection. Inspired by this analogy, we propose the VEN's sampling method Fig.~\ref{fig1}, which integrates generation operator and selection operator. The generation operator produces candidate configurations based on the current state, while the selection operator applies a selection strategy to choose more optimal configurations from these candidates. This generative and selective sampling strategy effectively avoids getting trapped in local minima, enhancing the diversity and comprehensiveness of the sampling process.

\begin{algorithm}[H]
\caption{\label{al} VEN's Sampling Method}
\begin{algorithmic}[1]
\Require Input configuration $\ensuremath{s \in \mathbb{R}^N}$, configuration size $N$, number of candidate configurations $K$, number of flipped sites $M$, probability calculation function $\ensuremath{G(\cdot)}$
\Ensure Output configuration $\ensuremath{s_{\text{out}} \in \mathbb{R}^N}$

\State Initialize an empty candidate array $\ensuremath{C \in \mathbb{R}^{K \times N}}$
\State Initialize an empty probability array $\ensuremath{P \in \mathbb{R}^K}$
\For{$i = 1$ to $K$}
    \State Randomly select a set $Q$ of $M$ unique positions from $\{1, 2, \ldots, N\}$
    \State Set $s' \gets s$ \Comment{Create a copy of \ensuremath{s}}
    \For{each position $q$ in $Q$}
        \State $s'[q] \gets 1 - s'[q]$ \Comment{Flip the value at position $q$}
    \EndFor
    \State $C[i] \gets s'$ \Comment{Store the modified configuration in \ensuremath{C}}
\EndFor

\For{$i = 1$ to $K$}
    \State $P[i] \gets G(C[i])$ \Comment{Calculate the probability using function \ensuremath{G}}
\EndFor

\State Normalize the probability array $P$: $P \gets \dfrac{P}{\sum_{i=1}^{K} P[i]}$ \Comment{Normalize the probabilities}

\State Randomly select an index $j$ based on the normalized probabilities in $P$
\State Set $s_{\text{out}} \gets C[j]$ \Comment{Set the output configuration}
\State \Return $s_{\text{out}}$ \Comment{Output the selected configuration}

\end{algorithmic}
\end{algorithm}

The generation operator, designed for models of size $L \times L$, begins with an initial configuration $\mathbf{s}$. It randomly selects $n$ positions—where $n$ is a hyperparameter—and flips the spins at these locations to generate a new candidate configuration $\mathbf{s'}^{(1)}$. This process is repeated to produce $M$ candidate configurations $\mathbf{s'}^{(1)}, \mathbf{s'}^{(2)}, \ldots, \mathbf{s'}^{(K)}$. The advantage of creating candidate configurations lies in its ability to increase sampling diversity under high-temperature conditions and to assist in escaping local minima at low temperatures, facilitating the exploration of more optimal solutions.

Following the generation of candidate configurations, the selection operator is responsible for determining the subsequent configuration. It first employs a neural network to compute the joint probability $p_\theta(\mathbf{s'}^{(i)})$ for each candidate configuration $\mathbf{s'}^{(i)}$. These joint probabilities are then normalized to obtain the selection probability for each candidate configuration:
\begin{align}
P(\mathbf{s'}^{(i)}) = \frac{p_\theta(\mathbf{s'}^{(i)})}{Z'},
\end{align}
where $Z' = \sum_{i=1}^K p_\theta(\mathbf{s'}^{(i)})$ serves as the normalization constant. By this approach, the selection operator adopts a strategy akin to a "roulette wheel" selection, choosing the next configuration based on the normalized probability distribution. This strategy ensures that configurations with higher probabilities—typically corresponding to lower energy states—are more likely to be selected, thereby accelerating the convergence towards the global optimal solution.

\section{Upper bound of the energy}
\label{sec:Upper}

The similarity between VEN and the process of evolutionary algorithms in searching for optimal solutions allows us to effectively overcome model limitations and find the ground state of energy. VEN gradually reaches the energy minimum through sampling. To analyze the effectiveness of the algorithm, we examine the maximum energy of the selected samples during the sampling process, which serves as an upper bound on the energy. If this upper bound decreases as the temperature is lowered, it indicates that the algorithm can learn the characteristics of the model, thereby effectively locating the energy minimum.

Inspired by the recent work on binary optimization~\cite{chen_monte_2023}, we would like to specifically analyze the upper bound of energy corresponding to this sampling method. Consider a binary system of $n$ lattice points $($such as the $1/2$ Ising spin$)$, which has a total of $N = 2^n$ spin configurations $\mathcal{B}_n = \{s_1, \ldots, s_N\}$, ordered by energy from low to high as $\text{E}(s_1) \leq \text{E}(s_2) \leq \cdots \leq \text{E}(s_N)$. A generation operator, given an input configuration $s_0$, generates a set of candidate spin configurations $\mathcal{N}(s_0) = \{s_1, s_2, s_3, \ldots, s_K\}$ with $\text{E}(s_1) \leq \text{E}(s_2) \leq \cdots \leq \text{E}(s_K)$.

Since the selection operator assigns selection weights according to the Boltzmann distribution, configurations with lower energy have a higher selection probability. When the temperature is sufficiently low, the configuration with the lowest energy will be chosen. Thus, in the following analysis, we directly select the configuration with the lowest energy:
\begin{align}
s = \arg\min_{s'} \text{E}(s') = T(s_0), s' \in \mathcal{N}(s_0).
\end{align}

\begin{figure}
\includegraphics[width=0.45\textwidth]{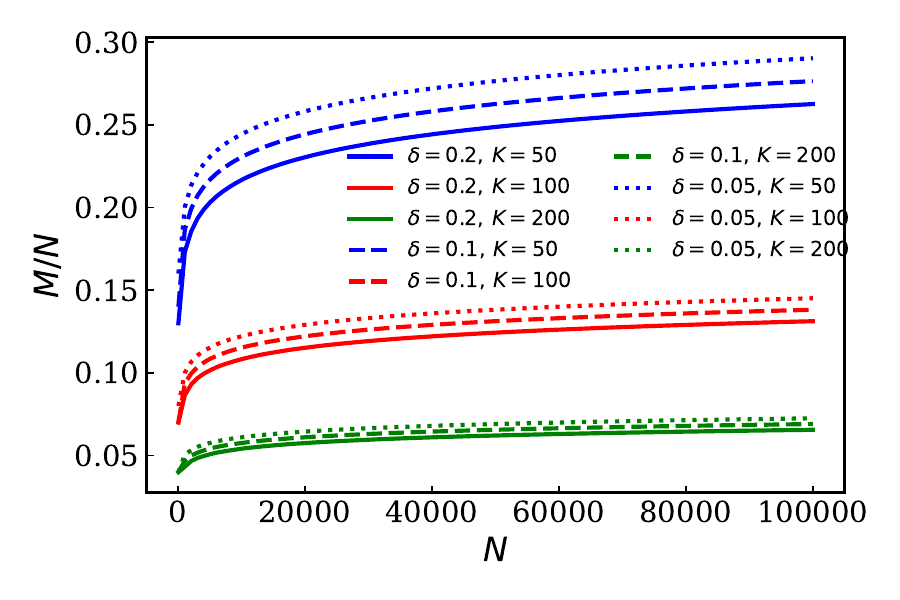} 
 \caption{\label{analy1}
The proportion by which the upper bound of the total number of configurations $N$ is compressed for a specific probability $\delta$ and number of generations $K$.}%
\end{figure}

Here, the operator $T$ is used to provide the final configuration through the generation operator and selection operator, given an initial configuration as an input. It can be shown that the upper bound of the energy of the spin configuration obtained through the $T$ operator satisfies with at least $1 - \delta$ probability $($See details in Appendix \ref{sec:proof}$)$:
\begin{align}
    \text{E}(T(s)) &\in [\text{E}(s_1), \text{E}(s_M)], \forall s \in \mathcal{B}_n, \label{upper-bound-1}
\end{align}
where $\delta \in (0, 1)$, $M = \left\lceil \frac{\log(N / \delta)}{K} N \right\rceil$. Eq.~(\ref{upper-bound-1}) demonstrates that through the operator $ T $, the energy range of the configuration will shrink, from $ [\text{E}(s_1), \text{E}(s_N)] $ to $ [\text{E}(s_1), \text{E}(s_M)] $. As shown in Fig.~\ref{analy1}, as the number of generated configurations $K$ increases, the upper bound of energy decreases more significantly, indicating a stronger correlation between the energy upper bound and $K$. Therefore, as long as $K$ is sufficiently large, the upper bound of energy can be reduced to a very low value with a high probability (with a small $\delta$).

Considering the joint probability of each configuration, the expected value of the energy through $T$ satisfies:
\begin{align}
    \mathbb{E}_{p_\theta}[\text{E}(T(x))] &\leq 
        \sum_{i=1}^{M-1} p_\theta(s_i) \text{E}(s_i)
        + \left(1 - \sum_{i=1}^{M-1} p_\theta(s_i)\right) \text{E}(s_M), \label{upper-bound-2}
\end{align}
where $\mathbb{E}_{p_\theta}[\text{E}(\cdot)]$ represents the expectation of the energy under the joint probability distribution $p_\theta$. Eq.~(\ref{upper-bound-2}) indicates that, after considering the joint probability of the configurations, configurations with lower energy are assigned higher weights, resulting in a lower expected value of the energy.

It should be emphasized that the generation operator and the selection operator significantly influence the energy range of configurations and the neural network’s control over the configuration probability weights, ultimately contribut to the reduction of the algorithm’s energy upper bound. Initially, the operator $T$ narrows the energy range of configurations, ensuring that the generated candidate configurations have lower energy values. Subsequently, the neural network is optimized by minimizing the KL divergence (Eq.~\ref{eq:KL}), aligning the predicted probability distribution with the true Boltzmann distribution, further reducing the energy range and accurately sampling the ground state configuration. 

\section{Numerical Demonstrations}
\label{sec:Num}

We employed the masked autoencoder for distribution estimation neural network~\cite{germain_made_2015} architecture, trained using the Adam optimizer. The input and output layers of the network have dimensions that are consistent with the size of the lattice model. The output layer utilizes a Sigmoid activation function, with the output at position $(i, j) \in (L, L)$ representing the probability $p_\theta(s_{ij})$ of the corresponding lattice site.

\subsection{2D Ising Model}

We first apply the VEN to the 2D ferromagnetic Ising model with a lattice size of $L \times L$ under periodic boundary conditions. The Hamiltonian reads:
\begin{align}
H = -J\sum_{\langle i,j \rangle}s_i s_j,
\end{align}
where $J=1$, and $<i,j>$ represents summation over all the nearest-neighbor sites.

\begin{figure}
\includegraphics[width=0.45\textwidth]{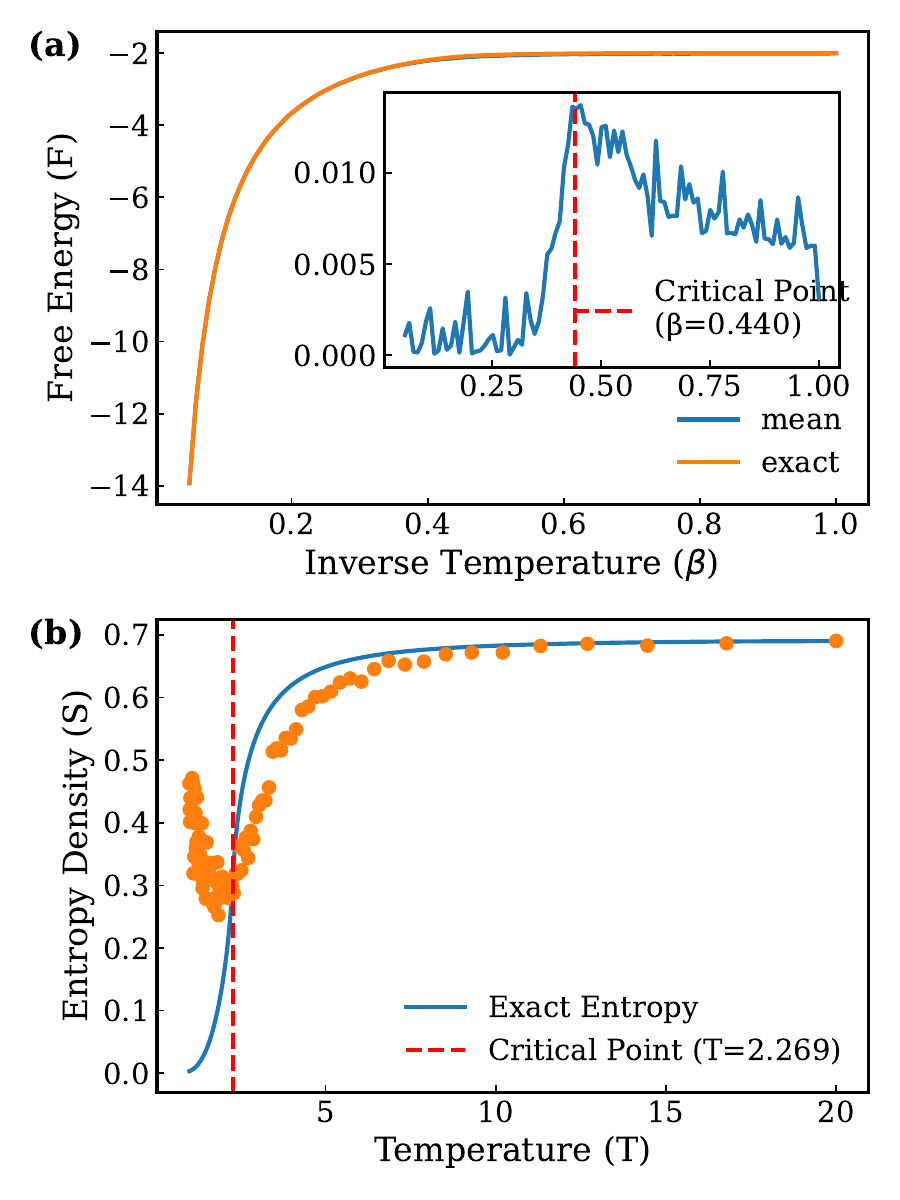} 
 \caption{\label{fig2} (a) The relationship between the free energy of the Ising model and the inverse temperature $\beta$, with the inset depicting the relative error between the simulated free energy and the exact free energy. (b) Illustrates the relationship between entropy and temperature $T$, where the blue line represents the exact solution. }%
\end{figure}

Free energy is a crucial thermodynamic quantity that directly reflects the equilibrium properties and dynamic behavior of the system. Therefore, we first analyzed the trend of free energy as a function of the inverse temperature $\beta$, as shown in Fig.~\ref{fig2}(a). During the simulation, $\beta$ was incrementally increased, and the simulated free energy trend closely matched the exact solution~\cite{onsager_crystal_1944, gabrie_adaptive_2022man_glassy_nodate}. The relative error was small for $\beta$ values below the critical point but increased significantly near the critical point and then gradually decreased as $\beta$ continued to rise. Additionally, due to the stochastic sampling used in the generation operator, some fluctuations in the relative error were observed.

Since both entropy and energy influence free energy, we further analyzed the entropy variation with temperature to investigate the reason for the peak in the relative error of free energy at the phase transition point, as shown in Fig.~\ref{fig2}(b). During the simulation, the temperature $T$ was gradually decreased. When the temperature was above the critical temperature, entropy decreased as temperature decreased; however, near the critical temperature, the trend reversed, with entropy increasing as the temperature decreased. This anomalous increase in entropy is the primary reason for the significant increase in the relative error of free energy. Although entropy continued to increase after the phase transition, its contribution to the free energy gradually diminished as the temperature decreased, leading to a reduction in the relative error of free energy. The anomalous increase in entropy can be attributed to the phase transition behavior of the model and the truncation in the VEN algorithm.

At the critical point, the Ising model transitions from short-range to long-range correlations, making it difficult for the neural network to capture these features quickly. As a result, the probability estimation $p_\theta(s_{ij})$ for certain lattice points may be inaccurate. Additionally, at the phase transition point, the Ising model undergoes symmetry breaking, where all spins align in the same direction. The presence of a single dominant configuration leads to overfitting of the neural network, causing it to lose generalization capability and eventually fail. Under these conditions, the neural network provides incorrect probability estimates for certain lattice points, and the overfitting prevents the network from making appropriate adjustments, resulting in an unexpected increase in entropy.

Another reason for the increase in entropy is the truncation strategy introduced in the algorithm. To avoid the selection probability of all candidate configurations being zero, a truncation value $P_{low}$ was introduced, ensuring that each configuration has a minimum selection probability, such that $P_{low} \leq P_{s'}$. At low temperatures, even if the ground state configuration is absent among the candidate configurations, the algorithm still selects a configuration for the next iteration. These incorrect configurations introduce biases in the statistical average, leading to an increase in entropy.

Although the anomalous increase in entropy at the critical point results in a peak in the relative error of free energy, entropy contributes relatively little to the free energy in the ground state ($T \to 0$). Under ground state conditions, the free energy is primarily determined by the lowest energy configuration. Next, we will consider the ground state of the model with quenched disorder to further demonstrate the efficiency of VEN in locating the ground state of the system.

\subsection{Sherrington-Kirkpatrick Model}

We further apply VEN to the classical Sherrington-Kirkpatrick (SK) model with $N$ lattice sites, which is a typical spin glass model~\cite{sherrington_solvable_1975}. The corresponding Hamiltonian is given as:
\begin{align}
H = -\sum_{i,j} J_{ij} s_i s_j,
\end{align}
where $ij$ represents the lattice positions, and the coupling coefficients $J_{ij}$ follow a Gaussian distribution with a mean of 0 and a variance of $1/N$. Note that all sites are interconnected.

\begin{figure}
\includegraphics[width=0.45\textwidth]{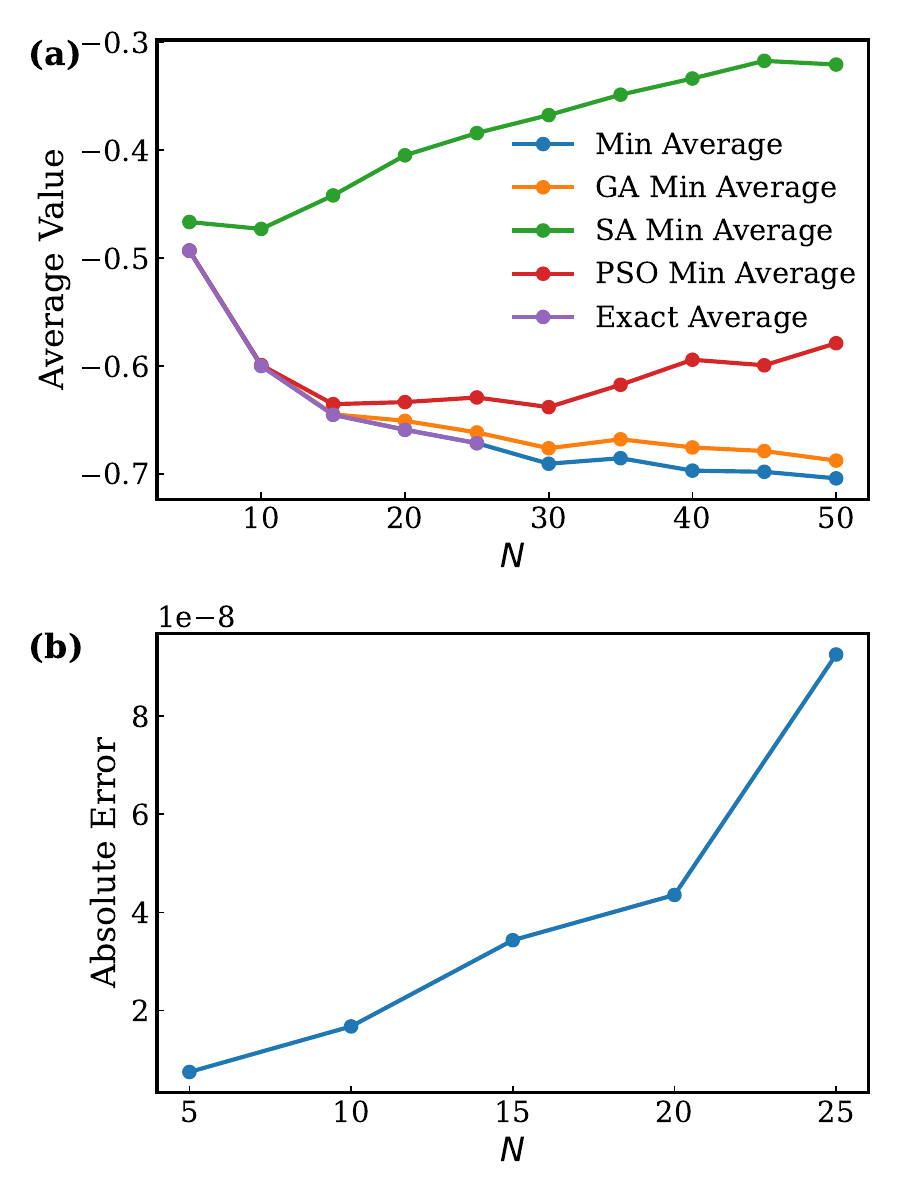} 
 \caption{\label{fig3}
(a) The relationship between the average ground state energy of the SK model obtained by different algorithms and the number of lattice sites.  GA, SA, and PSO represent the Genetic Algorithm, Simulated Annealing Algorithm, and Particle Swarm Optimization Algorithm. (b) The absolute error between the ground state energy obtained by VEN and the exact solution. }%
\end{figure}

\begin{figure}
\includegraphics[width=0.45\textwidth]{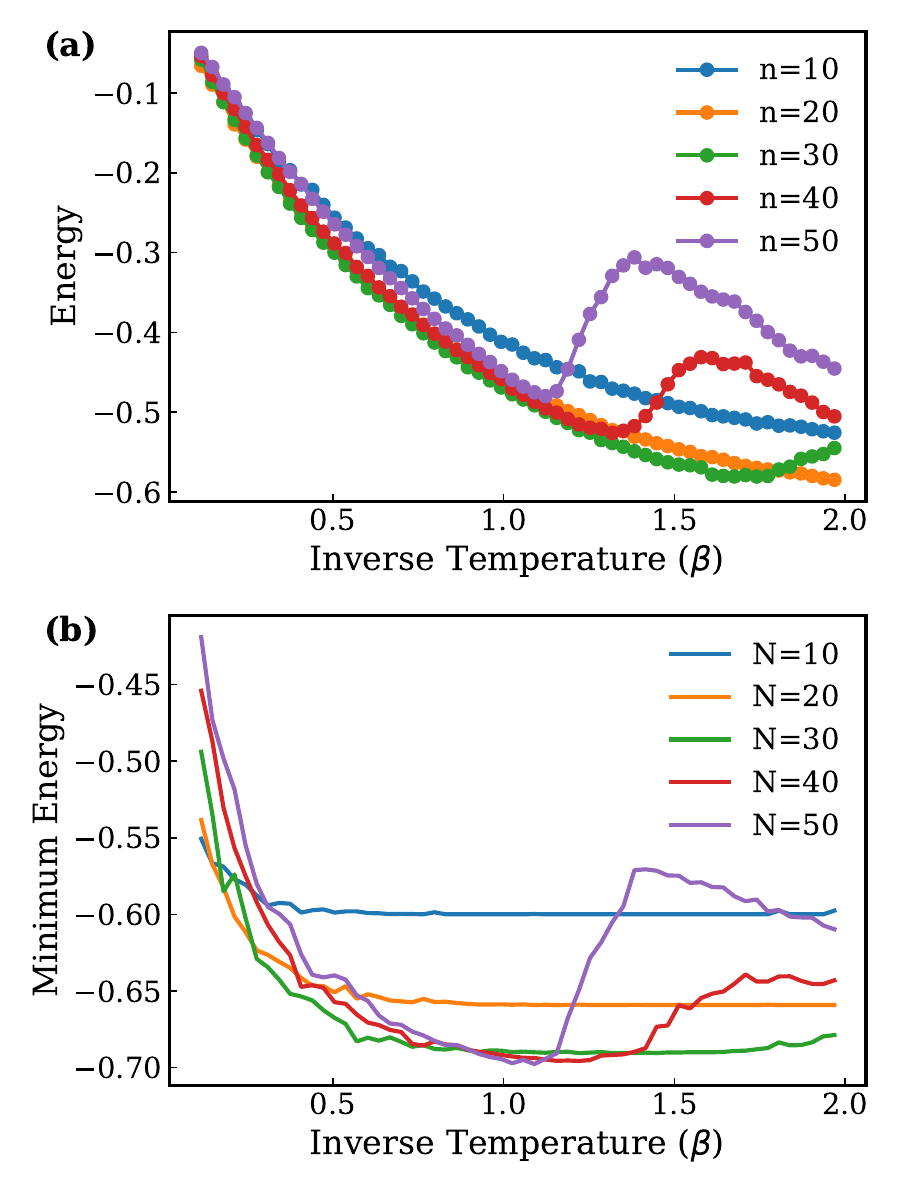} 
 \caption{\label{fig4}%
(a) The change in the average energy of candidate configurations with respect to $\beta$. (b) The change in the energy of candidate configurations with respect to $\beta$.
 }%
\end{figure}

The SK model plays a vital role in investigating disordered systems and spin glass behavior, and is widely used to describe complex combinatorial optimization problems. The ground state problem of the SK model is NP-hard due to the presence of significant frustration effects and complex interactions. It means that the complexity of solving the problem grows exponentially with the system size. Frustration leads to the existence of multiple local minima caused by inconsistent interactions in the system, making the solution process extremely challenging~\cite{pelikan_finding_2008}.

When $N$ is small, the exact solution of the SK model can be obtained, but as $N$ increases, the solution time grows exponentially, necessitating the use of approximate methods (e.g., simulated annealing and evolutionary algorithms). To evaluate the performance of different algorithms in finding the ground state of the SK model, we randomly generated 100 initial configurations for different lattice sizes $N$ and calculated the average ground state energy using various algorithms. Fig.~\ref{fig3}(a) shows a comparison between the exact solution, VEN, and other evolutionary algorithms, which were implemented using the scikit-opt library\cite{10.5555/1953048.2078195}. To clarify the relative performance of different algorithms in solving the SK model, we compared VEN with other evolutionary algorithms. As $N$ increases, the frustration effect becomes more pronounced. The simulated annealing algorithm tends to yield higher energy levels, indicating its tendency to get trapped in a local minima, while particle swarm optimization and genetic algorithms achieve relatively lower energy levels, suggesting their ability to overcome local minima. In particular, the genetic algorithm performs better due to its adaptability to discrete optimization problems. In contrast, the VEN algorithm yields the lowest average energy, and in Fig.~\ref{fig3}(b), the absolute error between VEN and the exact solution for small sizes is nearly zero, demonstrating the significant advantage of VEN in finding the ground state.

In order to further understand the process of VEN in solving the ground state energy, we studied its behavior during the gradual increase of the inverse temperature, as shown in Fig.~\ref{fig4}(a). It was observed that energy fluctuations gradually converge, and at the ground state, only upward energy fluctuations occur, indicating that the exploration region is gradually converging, and the average energy is also converging to the ground state. Additionally, the overall trend of the average energy is steadily decreasing, which is consistent with theoretical expectations. However, for larger lattice models, the average energy exhibits a sudden increase at a certain value of $\beta$, and the larger the lattice, the smaller the value of $\beta$ at which this increase occurs. This is due to the use of neural networks with the same depth and a fixed number of hidden neurons, and due to the limitations of the network's expressive capacity, larger lattice models are more prone to network failure, leading to an increase in energy.

We also analyzed the trend of the lowest energy among candidate configurations sampled by VEN as a function of $\beta$, as shown in Fig.~\ref{fig4}(b). It can be seen that the minimum energy initially drops sharply, exploring the lowest energy configuration. Despite sudden increases in energy during sampling due to the limited expressive capacity of the neural network, by recording the lowest energy point throughout the sampling process, we can still find a solution close to the ground state.

\section{Conclusions and Discussions}
\label{sec:Conclusions}
In summary, we proposed VEN, which integrates neural networks for variational free energy and evolutionary algorithms for sampling. 
By use of both the generation and selection operators, this sampling approach increases the number of samples, reduces the number of neural network calls, and allows for escaping local minima to find the ground state energy through random sampling. We also theoretically analyze how random point selection can successfully reduce the energy and give rise to a upper bound energy. 
Moreover, we found that VEN could successfully learn the phase transition point for Ising model and outperform usual optimization algorithms for the ground state energy for SK model.


The VEN algorithm may have two promising directions: further optimization of this sampling method and integration of more efficient neural networks. The construction of the generation  and selection operators here is relatively rudimentary; for specific problems, there is potential to further exploit marginal probabilities and develop more efficient operators. In fact, methods such as crossover operators from genetic algorithms could be considered. Regarding neural networks, combining more efficient networks could be much beneficial. In order to reduce variance, one could integrate the existing graph neural networks~\cite{schuetz_graph_2022}, MPVAN~\cite{ma_message_2024}. In addition, HVN~\cite{bialas_hierarchical_2022} could be considered to further optimize runtime.
%

\nocite{*}

\begin{acknowledgments}
The authors thank Tao Qin for useful discussions. This work was financially supported by the National Key R$\&$D Program of the MOST of China (Grant No. 2024YFA1611300), the National Natural Science Foundation of China under Grants (No. U2032164 and No. 12174394). J.Z. was also supported by HFIPS Director’s Fund (Grants No. YZJJQY202304 and No. BJPY2023B05), Anhui Provincial Major S$\&$T Project (s202305a12020005) and the Basic Research Program of the Chinese Academy of Sciences Based on Major Scientific Infrastructures (grant No. JZHKYPT-2021-08) and the High Magnetic Field Laboratory of Anhui Province under Contract No. AHHM-FX-2020-02.
\end{acknowledgments}

\appendix

\section{Appendix: Proof of existence of the upper bound}
\label{sec:proof}

In the Appendix, we follow the spirit ~\cite{chen_monte_2023} to prove the existence of the upper bound of ground state energy in spin models, that is, Eq.~(\ref{upper-bound-1}) and Eq.~(\ref{upper-bound-2}) in the main text.
\newline
(I) For Eq.~(\ref{upper-bound-1}), because the selection operator selects the candidate spin configuration with the lowest energy, for all $1 \leq i \leq N$, $\text{E}(T(s_i)) \leq \text{E}(s_{cand}), s_{cand} \in \mathcal{N}(s_i)$, it can be expressed as
$$\mathbb{P}\left(\text{E}\left(T\left(s_i\right)\right) > \text{E}\left(s_M\right)\right) \leq \mathbb{P}\left(\cup_{s_{cand}}\left\{\text{E}(s_{cand}) > \text{E}\left(s_M\right)\right\}\right),$$
where $s_{cand} \in \mathcal{N}\left(s_i\right)$.

If there exists a spin configuration $s_{cand}'$ in $\mathcal{N}(s_i)$ such that $\text{E}(s_{cand}') \leq \text{E}(s_M)$, we have 
$$\mathbb{P}\left(\cup_{s_{cand}}\left\{\text{E}(s_{cand}) > \text{E}\left(s_M\right)\right\}, s_{cand} \in \mathcal{N}(s_i)\right) = 0.$$ 
If there is no such configuration $s_{cand}'$, then 
$$\mathbb{P}\left(\cup_{s_{cand}}\left\{\text{E}(s_{cand}) > \text{E}\left(s_M\right)\right\}\right) = \prod_{s_{cand}} \mathbb{P}\left(\text{E}(s_{cand}) > \text{E}(s_M)\right).$$ 
Therefore, decompose the joint probability:
$$\mathbb{P}\left(\cup_s\left\{\text{E}(s) > \text{E}(s_M)\right\}\right) \leq \prod_s \mathbb{P}\left(\text{E}(s) > \text{E}(s_M)\right).$$ 
Considering $s$ as a randomly selected configuration from $\mathcal{B}_n$, $\mathbb{P}\left(\text{E}(s) > \text{E}(s_M)\right) = (N - M) / N$,
$$\mathbb{P}\left(\text{E}\left(T\left(s_i\right)\right) > \text{E}\left(s_M\right)\right) \leq \left(1 - \frac{M}{N}\right)^K$$

Combining the inequality $1 - x \leq e^{-x}$ with the definition of $M$, we get:
$$\mathbb{P}\left(\text{E}\left(T\left(s_i\right)\right) > \text{E}\left(s_M\right)\right) \leq \exp\left(-\frac{M}{N}K\right) \leq \frac{\delta}{N}$$

Finally, the relation is:
\begin{align*}
\mathbb{P}\left(\text{E}(T(x)) \leq \text{E}(s_M)\right) &\geq 1 - \sum_{i=1}^N \mathbb{P}\left(\text{E}\left(T\left(s_i\right)\right) > \text{E}(s_M)\right) \\
&\geq 1 - \delta.
\end{align*}
\newline
(II) For Eq.~(\ref{upper-bound-2}), for $M \leq i \leq N$, we have the relation:
$$\mathbb{P}\left(\text{E}\left(T\left(s_i\right)\right) > \text{E}(s_M)\right) \leq \frac{\delta}{N}$$
For the operator $T$, with a probability of $1 - \delta$:
$$\text{E}\left(T\left(s_i\right)\right) \leq \text{E}(s_M), \quad \forall M \leq i \leq N.$$
Thus:
\begin{align*}
&\mathbb{E}_{p_\theta}[\text{E}(T(x))] \\
&= \sum_{i=1}^N p_\theta(s_i) \text{E}(T(s_i)) \\
&\leq \sum_{i=1}^{M-1} p_\theta(s_i) \text{E}(s_i) + \left(1 - \sum_{i=1}^{M-1} p_\theta(s_i)\right) \text{E}(s_M).
\end{align*}
This completes the proof.

\bibliography{VEN}

\end{document}